\documentclass[aps,prl,twocolumn, reprint,superscriptaddress]{revtex4}
\usepackage{graphicx}

\begin{document}

\title{Long range dipole-dipole interaction in atomic vapors probed by double-quantum two-dimensional coherent spectroscopy}


\author{Shaogang Yu}
\affiliation
{Department of Physics, Florida International University, Miami, Florida 33199, USA}

\affiliation
{State Key Laboratory of Magnetic Resonance and Atomic and Molecular Physics, Wuhan Institute of Physics and Mathematics, Chinese Academy of Sciences, Wuhan 430071, China}

\affiliation
{University of Chinese Academy of Sciences, Beijing 100049, China}

\author{Michael Titze}
\affiliation
{Department of Physics, Florida International University, Miami, Florida 33199, USA}

\author{Yifu Zhu}
\affiliation
{Department of Physics, Florida International University, Miami, Florida 33199, USA}

\author{Xiaojun Liu}

\affiliation
{State Key Laboratory of Magnetic Resonance and Atomic and Molecular Physics, Wuhan Institute of Physics and Mathematics, Chinese Academy of Sciences, Wuhan 430071, China}

\author{Hebin Li}
\email[Email address: ]{hebin.li@fiu.edu}
\affiliation
{Department of Physics, Florida International University, Miami, Florida 33199, USA}

\begin{abstract}
Optical double-quantum two-dimensional coherent spectroscopy (2DCS) was implemented to probe interatomic dipole-dipole interactions in both potassium and rubidium atomic vapors. The dipole-dipole interaction was detected at densities of $4.81 \times 10^8$ cm$^{-3}$ and $8.40 \times 10^9$ cm$^{-3}$ for potassium and rubidium, respectively, corresponding to a mean interatomic separation of 15.8 $\mu$m or $3.0\times 10^5a_0$ for potassium and 6.1 $\mu$m or $1.2\times 10^5a_0$ for rubidium, where $a_0$ is the Bohr radius. We report the lowest atomic density at which dipole-dipole interactions are detected. The experimental results confirm the long range nature of the dipole-dipole interaction which is critical for understanding many-body physics in atoms/molecules. The long range interaction also has implications in atom-based applications involving many-body interactions. Additionally, we demonstrated that double-quantum 2DCS is sufficiently sensitive to probe dipole-dipole interaction at densities that can be achieved with cold atom in a magneto-optical trap, paving the way for double-quantum 2DCS studies of cold atoms and molecules. The method can also open a new avenue to study long-range interactions in solid states systems such as quantum dots and color centers in diamonds. 
\end{abstract}

\maketitle

Neutral atoms without permanent dipole moments can interact due to transition-induced dipole moments \cite{King1939}. This dipole-dipole interaction introduces important many-body effects in systems such as cold atoms/molecules \cite{Weiner1989,Jones2006}, optical atomic clocks \cite{Ludlow2008,Swallows2011}, atomic vapors \cite{Maki1991,Sautenkov1996,VanKampen1997,Li2008,Li2009,Shen2007b,Eden2008,Cundiff2002a,Lorenz2005a,
Lorenz2008a,Dai2012,Gao2016,PhysRevLett.120.233401}, and photosynthesis \cite{Engel2007}. For instance, the interaction between atoms trapped in an optical lattice affects the precision of an optical atomic clock \cite{Swallows2011}; the interaction is essential in atom-based quantum simulators \cite{Bernien2017,Mazurenko2017,Trotzky2012} to enable many-body quantum simulation. In many cases, the interatomic distance can extend to tens of $\mu$m. Is the dipole-dipole interaction still effective at such a long range relative to the size of an atom? Some theories treating dipole-dipole interactions in cold/ultracold atoms \cite{Weiner1989} and atomic vapors \cite{Szudy1975,Reuven1975,Smith1973,Ali1965,Anderson1949,Lewis1980} only account for binary interactions at short ranges at the order of Weisskopf radius, the impact parameter that gives a unity optical phase shift \cite{Weisskopf1932, Thorne1999}. On the other hand, Leegwater and Mukamel \cite{Leegwater1994a} suggested that the interaction should be considered at any density, based on their exciton model of atomic vapors that includes many-body dipole-dipole interactions at all ranges. Therefore, it is critical to experimentally confirm the effective interaction range in understanding the role of dipole-dipole interactions.

In this letter, we report the experimental detection of 10-$\mu$m range dipole-dipole interactions in atomic vapors. Double-quantum optical two-dimensional coherent spectroscopy (2DCS) measurements were performed on both potassium (K) and rubidium (Rb) atomic vapors using a collinear 2DCS setup. The spectra show signals due to dipole-dipole interactions in atomic vapors at room temperature with atomic densities as low as  $4.81 \times 10^8$ cm$^{-3}$ and $8.40 \times 10^9$ cm$^{-3}$ for K and Rb, respectively. At these densities, the mean atomic separation $\langle R\rangle =2(\frac{4\pi}{3}N)^{-1/3}$ is 15.8 $\mu$m (or $3.0\times10^5a_0$) for K and 6.1 $\mu$m (or $1.2\times10^5a_0$) for Rb, where $a_0$ is the Bohr radius. A simulation based on the optical Bloch equations can reproduce the experiment spectra, revealing effects of dipole-dipole interaction. Compared to previous reports, the lowest density at which the interaction was detected in our experiment is at least 3 orders of magnitude lower than the lowest densities previously measured by other techniques. It is estimated that 227 atoms are measured in the experiment, confirming the technique's sensitivity of detecting dipole-dipole interactions. By integrating with a femtosecond frequency comb \cite{PhysRevLett.120.233401,Lomsadze2017b}, this technique can be developed to provide quantitative information about many-body dipole-dipole interactions in atomic vapors and cold atoms. The capability to detect long-range interactions can open a new avenue to study many-body interactions in solid state systems such as quantum dots and color centers in diamonds.

Various spectroscopic techniques have been used to study dipole-dipole interactions in atomic ensembles. Photoassociation spectroscopy \cite{Weiner1989} in cold atoms can map out the potential energy curve to extract the information about interaction. For hot atomic vapors, early studies primarily focused on very dense Rb and K atomic vapors with densities $N>10^{15}$ cm$^{-3}$. In the frequency domain, selective reflection spectra \cite{VanKampen1997,Li2009,Li2008,Maki1991,Sautenkov1996} exhibit line shift and broadening due to interactions. In the time domain, quantum beating experiments \cite{Shen2007b,Eden2008} measured the energy shift of $d$ states in Rb atoms and a quantitative explanation of the results requires to account for interactions among at least five atoms; transient four-wave mixing (TFWM) spectroscopy \cite{Cundiff2002a,Lorenz2005a,Lorenz2008a} can probe the transient dynamics of interactions to reveal non-Markovian behaviors. However, the effects of dipole-dipole interactions on the spectra in both frequency and time domains become more subtle, especially in the presence of thermal motion, as the atomic density decreases. The above-mentioned techniques often have difficulties in detecting dipole-dipole interactions at atomic densities lower than $10^{15}$ cm$^{-3}$.

Optical double-quantum 2DCS has been demonstrated as a sensitive probe of dipole-dipole interactions in atomic vapors \cite{Dai2012,Gao2016,PhysRevLett.120.233401}. In double-quantum 2DCS, the signal is a result of the double-quantum coherence between the ground state and a doubly excited state that can be either a single atom state or a collective state of two atoms. In the latter case, the contributions of all possible excitation pathways cancel out if the two atoms do not interact \cite{Gao2016}. In the presence of an interaction, the resulting energy shift and difference in dephasing rates break the symmetry so that the cancellation is incomplete, giving rise to a nonzero double-quantum signal \cite{Gao2016}. Since a slight asymmetry can lead to an incomplete cancellation, double-quantum 2DCS provides an extremely sensitive and background-free detection to weak dipole-dipole interactions. However, previous measurements \cite{Dai2012,Gao2016,PhysRevLett.120.233401} were performed at atomic densities of $N>10^{12}$ cm$^{-3}$ in heated vapor cells. It was not clear if the dipole-dipole interaction can still be detected at the lowest density that can be conveniently achieved with a vapor cell at room temperature. Additionally, validating this technique at a typical density ($N\sim 10^{10}$ cm$^{-3}$) of cold atoms is necessary before implementing 2DCS on cold atoms in a magneto-optical trap (MOT).

\begin{figure}[bth]
\centering
  \includegraphics[width=1.0\columnwidth]{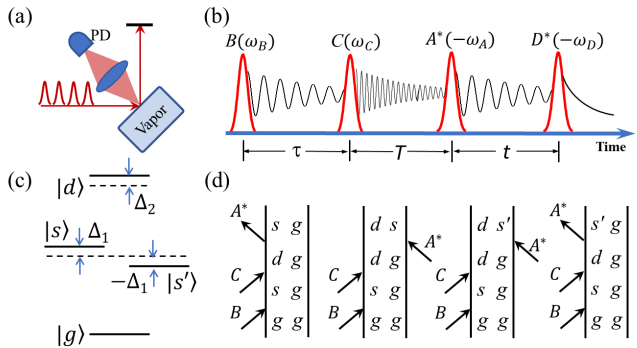}
  \caption{(a) Experimental schematic. (b) Time ordering of the excitation pulse sequence. (c) Energy diagram of two two-level atoms. (d) Four of the eight excitation pathways contributing to the double-quantum signal.}
  \label{fig:figure1}
\end{figure}

The same basic principle of using double-quantum 2DCS to probe dipole-dipole interactions, as previously reported \cite{Dai2012,Gao2016}, is also used in this study. However, the implementation is based on the collinear 2DCS \cite{Nardin2013,Tekavec2007} instead of the box geometry. As shown in Fig. \ref{fig:figure1}(a), four excitation pulses are incident on the window of a vapor cell and the fluorescence signal is recorded by a photodiode (PD). The time ordering of the four pulses, $B$, $C$, $A^*$, and $D^*$ is shown in Fig. \ref{fig:figure1}(b). Each pulse is phase modulated by an acousto-optic modulator (AOM) so that the pulses are tagged with different modulation frequencies ($\omega_B$, $\omega_C$, $-\omega_A$, and $-\omega_D$), where the negative sign corresponds to a conjugated pulse. To measure the double-quantum signal, the output from the PD is analyzed by a lock-in amplifier at the reference frequency $\omega_S=\omega_B+\omega_C-\omega_A-\omega_D$. The signal is measured as time delays $T$ and $t$ are scanned and then is Fourier transformed in both dimensions to generate a double-quantum 2D spectrum. Higher-order-quantum signals were also measured \cite{Bruder2015,Yu2018} in a similar setup by using same pulses multiple times and detecting at the proper reference frequency.

The collective states of two identical two-level atoms are shown in Fig. \ref{fig:figure1}(c), where $|g\rangle$ is the ground state with both atoms in the ground state, the singly excited states $|s\rangle$ and $|s'\rangle$ each have one atom in the excited state and the other in the ground state, and the doubly excited state $|d\rangle$ has both atoms in the excited state. We are interested in the processes that involve the double-quantum coherence between $|d\rangle$ and $|g\rangle$. Under the excitation of the pulse sequence in Fig. \ref{fig:figure1}(b), the first two pulses generate a double-quantum coherence and the third pulse converts the double-quantum coherence into a single-quantum coherence. There are eight possible excitation pathways. Double-sided Feynman diagrams in Fig. \ref{fig:figure1}(d) represent four pathways and switching $s$ and $s'$ gives the other four. To detect the resulting single-quantum coherence, the forth pulse converts single-quantum coherence into populations which emit fluorescence. We can select the fluorescence resulted from this particular excitation sequence by detecting the signal at the reference frequency $\omega_S$ via lock-in detection.  If the two atoms are independent, the four-level energy scheme in Fig. \ref{fig:figure1}(c) is symmetric with $\Delta_{1,2}=0$, all eight pathways cancel out exactly resulting in no signal. However, the interaction between the two atoms introduces energy shifts ($\Delta_{1,2}\neq 0$) and a difference in the dephasing rates for the upper and lower transitions, leading to a nonzero signal in double-quantum 2D spectra. A theoretical model based on the exciton picture \cite{Dai2012} attributed the signal to interatomic dipole-dipole interactions.

\begin{figure}[bth]
\centering
  \includegraphics[width=1.0\columnwidth]{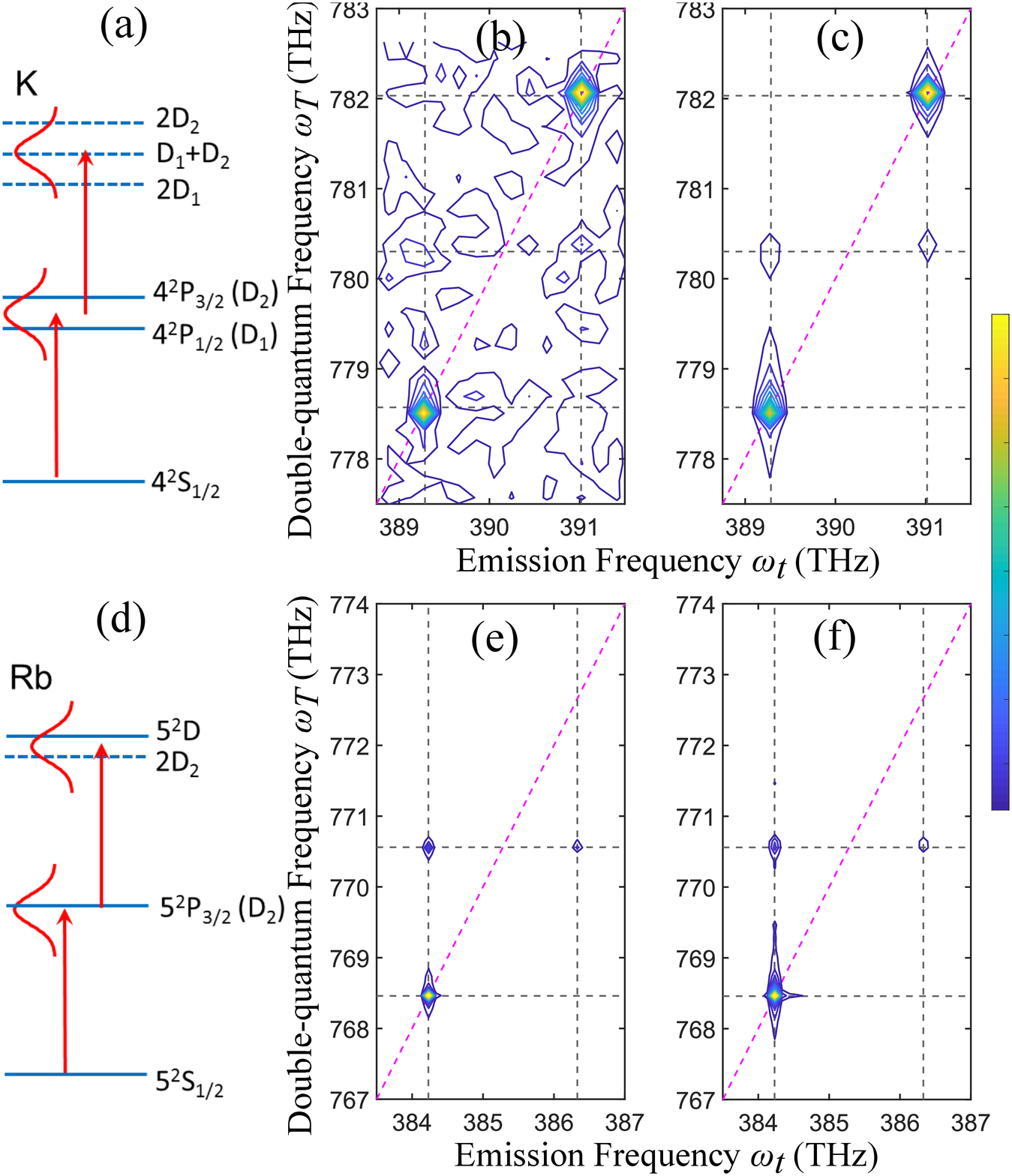}
  \caption{(a) Relevant energy levels of single (solid lines) and two (dashed lines) K atoms. Double-quantum 2D spectra of a K vapor at densities (b) $N=4.81\times 10^8$ cm$^{-3}$ and (c) $N=1.12\times 10^{10}$ cm$^{-3}$. (d) Relevant energy levels of single (solid lines) and two (dashed lines) Rb atoms. Double-quantum 2D spectra of a Rb vapor at densities (e) $N=8.40\times 10^9$ cm$^{-3}$ and (f) $N=2.29\times 10^{10}$ cm$^{-3}$.}
  \label{fig:figure2}
\end{figure}

Double-quantum 2DCS was performed to detect dipole-dipole interactions in both a K vapor and an Rb vapor each contained in a glass cell. With residue metal in the cells, the saturation vapor pressure can be estimated \cite{Nesmeyanov1963} from the temperature at the cold point and then is used to calculate the atomic density. Previous experiments \cite{Dai2012,Gao2016,PhysRevLett.120.233401} were done at elevated temperatures of 100 $^\circ$C or higher to reach atomic densities higher than $10^{12}$ cm$^{-3}$. In our current experiment, double-quantum signals resulting from dipole-dipole interactions were detected at room temperature (25 $^\circ$C) at which the atomic density is $N=4.81\times 10^8$ cm$^{-3}$ for K and $N=8.40\times 10^9$ cm$^{-3}$ for Rb. We also confirmed the signal at typical cold-atom densities of $\sim10^{10}$ cm$^{-3}$ in a MOT.

Double-quantum 2D spectra of K are shown in Figs. \ref{fig:figure2}(b) and (c) for $N=4.81\times 10^8$ cm$^{-3}$ and $N=1.12\times 10^{10}$ cm$^{-3}$, respectively. The emission frequency $\omega_t$ and the double-quantum frequency $\omega_T$ correspond to the Fourier transform of dynamics during time delays $t$ and $\tau$, respectively. For K, there are three doubly excited states $2D_1$, $D_1+D_2$, and $2D_2$ that are all two-atom states, as shown in Fig. \ref{fig:figure2}(a). These doubly excited states result in four peaks in the double-quantum 2D spectrum in Fig. \ref{fig:figure2}(c) with two on the diagonal line $\omega_T=2\omega_t$ and two off-diagonal. The off-diagonal peaks have a lower amplitude due to a weaker interaction and their amplitude is below the noise level in the 2D spectrum at the lower density. For Rb, the doubly excited states consist of a single-atom state $5^2D$ and a two-atom state $2D_2$, as shown in Fig. \ref{fig:figure2}(d). Double-quantum 2D spectra of Rb are shown in Figs. \ref{fig:figure2}(e) and (f) for $N=8.40\times 10^9$ cm$^{-3}$ and $N=2.29\times 10^{10}$ cm$^{-3}$, respectively. The diagonal peak is a result of the two-atom state while the off-diagonal peaks are contributed by the single-atom state. The 2D spectra at both densities have a sufficiently large signal-to-noise ratio to resolve all three peaks. These double-quantum 2D spectra are similar to the previously reported measurements at higher atomic densities ($>10^{12}$ cm$^{-3}$) for K \cite{Dai2012} and Rb \cite{Gao2016}.

\begin{figure}[hbt]
\centering
  \includegraphics[width=\columnwidth]{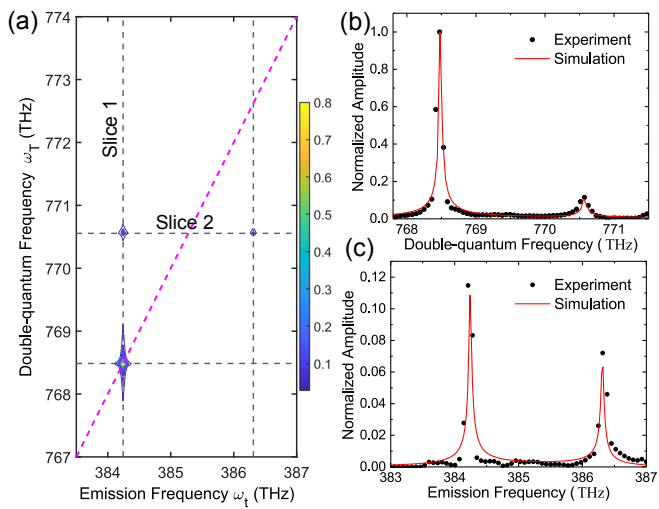}
  \caption{(a) Simulated double-quantum 2D spectrum of Rb. Comparison of slices taken from experimental and simulated 2D spectra along the directions denoted as (b) Slice 1 and (c) Slice 2. The black dots are experimental data and the red lines are simulation.}
  \label{fig:figure3}
\end{figure}

Based on the double-quantum excitation process illustrated in Fig. \ref{fig:figure1}, experimental 2D spectra in Fig. \ref{fig:figure2} can be reproduced by a simulation using the optical Bloch equations. The amplitude of the two-atom signal depends on energy shifts and changes in dephasing rates due to the interaction. For K, the double-quantum 2D spectra alone do not provide a sufficient constraint to determine the interaction strength. In the case of Rb, the 2D spectra include double-quantum signals from both the two-atom and the single-atom states. The relative amplitude of the two-atom signal compared to the single-atom signal provides an extra constraint to determine the energy shifts due to the interaction. Following the formalism in ref. \cite{Gao2016}, the double-quantum 2D spectrum of Rb can be calculated as
\begin{widetext}
\begin{eqnarray}
P^{(3)}(\omega_t,\omega_T)&=&\frac{S_0 \mu_{10}^2\mu_{21}^2 }{\omega_T-\omega_{20} + i\Gamma_{20}} (\frac{1}{\omega_t-\omega_{10}+ i\Gamma_{10}}-\frac{1}{\omega_t-\omega_{21}+ i\Gamma_{21}})+  \nonumber
\\
&&\frac{4S_0 \mu_{10}^4}{\omega_T-\omega_{dg} + i\Gamma_{dg}}(\frac{1}{\omega_t-\omega_{sg}+ i\Gamma_{sg}}-\frac{1}{\omega_t-\omega_{ds}+ i\Gamma_{ds}}-\frac{1}{\omega_t-\omega_{ds'}+ i\Gamma_{ds'}}+\frac{1}{\omega_t-\omega_{s'g}+ i\Gamma_{s'g}}), \label{eq:eq1}
\end{eqnarray}
\end{widetext}
where $\mu_{ij}$ are the dipole moments, $\Gamma_{ij}$ are the relaxation matrix elements, and $S_0=\frac{iNE_AE_BE_C}{16\pi\hbar^3}$ with $N$ being the atomic density, $E_{A,B,C}$ the electric field amplitudes of pulses, and $\hbar$ the reduced Planck constant. The first term is the contribution from single atom states which consist of ground state $|0\rangle = |5^2S_{1/2}\rangle$, singly excited state $|1\rangle= |5^2P_{3/2}\rangle$, and doubly excited state $|2\rangle = |5^2D\rangle$. The second term is the signal from two-atom states in Fig. \ref{fig:figure1}(c), where $\omega_{sg}=\omega_{10}+\Delta_1$, $\omega_{ds'}=\omega_{10}+\Delta_1+\Delta_2$, $\omega_{s'g}=\omega_{10}-\Delta_1$, $\omega_{ds}=\omega_{10}-\Delta_1+\Delta_2$ and $\omega_{dg}=2\omega_{10}+\Delta_2$. To produce the experimental spectrum in Fig. \ref{fig:figure2}(e), parameters are chosen to match both lineshapes and relative amplitudes. The simulated double-quantum 2D spectrum is shown in Fig. \ref{fig:figure3}(a). Two slices along directions labelled "Slice 1" and "Slice 2" are taken to compare with the same slices from the experimental spectrum. The slices are shown in Figs. \ref{fig:figure3}(b) and (c) for Slice 1 and 2, respectively, where the simulation fits the experimental data in both linewidth and relative amplitude. The parameters used in the simulations are $\Gamma_{10}=\Gamma_{21}=\Gamma_{sg}=\Gamma_{s'g}=130$ GHz, $\Gamma_{20}=150$ GHz, $\Gamma_{dg}=90$ GHz, $\Gamma_{ds}=\Gamma_{ds'}=135$ GHz. At the atomic density of $N=8.40\times 10^9$ cm$^{-3}$, the dipole-dipole interaction induces a 5-GHz difference between $\Gamma_{ds,ds'}$ and $\Gamma_{sg,s'g}$. The induced energy shifts $\Delta_{1,2}$ are estimated \cite{Cline1994} to be smaller than 1 KHz and thus negligible in equation (\ref{eq:eq1}). Therefore, at low densities, the primary effect of long-range dipole-dipole interactions is the change in dephasing rates, which contributes to the double-quantum signal from two-atom states.

\begin{figure}[bth]
\centering
  \includegraphics[width=1.0\columnwidth]{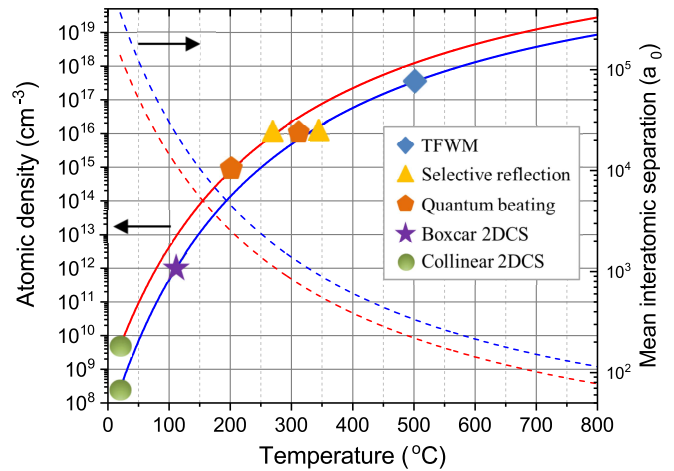}
  \caption{Sensitivity of detecting dipole-dipole interaction in atomic vapors by various techniques. Atomic density (solid lines) and mean interatomic separation (dashed lines) at different temperatures are shown for K (blue) and Rb (orange) vapors. The reported lowest density is marked with diamond for TFWM, triangle for selective reflection, pentagon for quantum beating, star for boxcar 2DCS, and circle for collinear 2DCS.}
  \label{fig:figure4}
\end{figure}

Several other spectroscopic techniques \cite{Maki1991,Sautenkov1996,VanKampen1997,Li2008,Li2009,Shen2007b,Eden2008,Cundiff2002a,Lorenz2005a,
Lorenz2008a,Dai2012,Gao2016,PhysRevLett.120.233401} have been used to probe the effects of dipole-dipole interactions in K and Rb atomic vapors. The effective range of dipole-dipole interactions revealed by these experiments varies substantially due to the variation in detection sensitivity of these methods. To illustrate the detected interaction range reported for different methods, the atomic density (solid lines) and the corresponding mean interatomic separation (dashed lines) are plotted as a function of temperature for K (blue) and Rb (orange) vapors in Fig. \ref{fig:figure4}. The reported lowest densities at which the dipole-dipole interaction was detected are marked with diamond for TFWM \cite{Cundiff2002a,Lorenz2005a,Lorenz2008a}, triangle for selective reflection \cite{Maki1991,Sautenkov1996,VanKampen1997,Li2008,Li2009}, pentagon for quantum beating \cite{Shen2007b,Eden2008}, and star for 2DCS in the boxcar geometry \cite{Dai2012,Gao2016}. The boxcar 2DCS was the most sensitive and can detect dipole-dipole interactions at a density of $10^{12}$ cm$^{-3}$ which is at least 3 orders of magnitude lower than the lowest densities for other methods. Using the collinear 2DCS technique, our experiment detects dipole-dipole interactions at a density of $4.81\times10^8$ cm$^{-3}$ for K and $8.40\times10^9$ cm$^{-3}$ for Rb, about 3 orders of magnitude lower than the previous limit.

Our measurements at low densities, together with previous work at high densities, indicate that dipole-dipole interactions should be considered at all densities ranging from $10^8$ to $10^{19}$ cm$^{-3}$. These account for a large range across 11 orders of magnitude in density for atomic vapors at temperatures from 25 to 800 $^\circ$C. Moreover, at the lowest density, the mean interatomic separation $\langle R \rangle=2(\frac{4\pi}{3}N)^{-1/3}$ is 15.8 $\mu$m or $3.0\times 10^5a_0$ for K and 6.1 $\mu$m or $1.2\times 10^5a_0$ for Rb. The result experimentally confirms the long range nature of dipole-dipole interaction with an effective interaction range up to 15.8 $\mu$m. This 10-$\mu$m range interaction might have implications in experiments with optical lattices since the interaction is not just limited to the same or nearest lattice sites but also can extend to the sites further away. The long-range interaction is crucial for atom-based quantum simulators \cite{Bernien2017,Mazurenko2017,Trotzky2012} to perform many-body simulation and the dipole-dipole interaction might provide an alternative approach. We also note that the excitation laser beam is focused onto the cell window with a beam waist of 10 $\mu$m and a Rayleigh range of 1.5 mm. At the lowest density ($4.81\times 10^8$ cm$^{-3}$) for K, the number of atoms in this cylindrical volume is estimated to be only 227, demonstrating the detection sensitivity of the technique. Two major technical concerns of implementing 2DCS in cold atoms are sensitivity and frequency resolution. Our experiment shows that collinear 2DCS has sufficient sensitivity for the number of atoms and the density in a typical MOT. The required frequency resolution can be achieved by using femtosecond frequency combs \cite{PhysRevLett.120.233401,Lomsadze2017b} in 2DCS.

In summary, the long-range dipole-dipole interaction is detected in both K and Rb atomic vapors by using double-quantum 2DCS. The collinear 2DCS provides a powerful technique to detect dipole-dipole interactions at an atomic density that is at least 3 orders of magnitude lower than the lowest density for other techniques. Based on the mean interatomic separation at the lowest densities in our experiment, the effective interaction range extends to over 10 $\mu$m or $10^5a_0$. The result confirms the long range nature of dipole-dipole interactions which should be considered in a large range of densities over 11 orders of magnitude, if not all densities \cite{Leegwater1994a}. The long range dipole-dipole interaction might also have important implications in experiments with optical lattices and atom-based quantum simulators, where many-body interaction plays a key role. Our experiment also addresses the sensitivity issue of performing double-quantum 2DCS in cold atoms/molecules, demonstrating the potential of 2DCS studies on many-body interactions and dynamics in cold atoms and molecules.


\begin{acknowledgments}
This material is based upon work supported by the National Science Foundation under Grant No. PHY-1707364.
\end{acknowledgments}


\begin{thebibliography}{38}
\expandafter\ifx\csname natexlab\endcsname\relax\def\natexlab#1{#1}\fi
\expandafter\ifx\csname bibnamefont\endcsname\relax
  \def\bibnamefont#1{#1}\fi
\expandafter\ifx\csname bibfnamefont\endcsname\relax
  \def\bibfnamefont#1{#1}\fi
\expandafter\ifx\csname citenamefont\endcsname\relax
  \def\citenamefont#1{#1}\fi
\expandafter\ifx\csname url\endcsname\relax
  \def\url#1{\texttt{#1}}\fi
\expandafter\ifx\csname urlprefix\endcsname\relax\def\urlprefix{URL }\fi
\providecommand{\bibinfo}[2]{#2}
\providecommand{\eprint}[2][]{\url{#2}}

\bibitem[{\citenamefont{King and van Vleck}(1939)}]{King1939}
\bibinfo{author}{\bibfnamefont{G.~W.} \bibnamefont{King}} \bibnamefont{and}
  \bibinfo{author}{\bibfnamefont{J.~H.} \bibnamefont{van Vleck}},
  \bibinfo{journal}{Phys. Rev.} \textbf{\bibinfo{volume}{55}},
  \bibinfo{pages}{1165} (\bibinfo{year}{1939}).

\bibitem[{\citenamefont{Weiner et~al.}(1999)\citenamefont{Weiner, Bagnato,
  Zilio, and Julienne}}]{Weiner1989}
\bibinfo{author}{\bibfnamefont{J.}~\bibnamefont{Weiner}},
  \bibinfo{author}{\bibfnamefont{V.~S.} \bibnamefont{Bagnato}},
  \bibinfo{author}{\bibfnamefont{S.}~\bibnamefont{Zilio}}, \bibnamefont{and}
  \bibinfo{author}{\bibfnamefont{P.~S.} \bibnamefont{Julienne}},
  \bibinfo{journal}{Rev. Mod. Phys.} \textbf{\bibinfo{volume}{71}},
  \bibinfo{pages}{1} (\bibinfo{year}{1999}).

\bibitem[{\citenamefont{Jones et~al.}(2006)\citenamefont{Jones, Tiesinga, Lett,
  and Julienne}}]{Jones2006}
\bibinfo{author}{\bibfnamefont{K.~M.} \bibnamefont{Jones}},
  \bibinfo{author}{\bibfnamefont{E.}~\bibnamefont{Tiesinga}},
  \bibinfo{author}{\bibfnamefont{P.~D.} \bibnamefont{Lett}}, \bibnamefont{and}
  \bibinfo{author}{\bibfnamefont{P.~S.} \bibnamefont{Julienne}},
  \bibinfo{journal}{Rev. Mod. Phys.} \textbf{\bibinfo{volume}{78}},
  \bibinfo{pages}{483} (\bibinfo{year}{2006}).

\bibitem[{\citenamefont{Ludlow et~al.}(2008)\citenamefont{Ludlow, Zelevinsky,
  Campbell, Blatt, Boyd, de~Miranda, Martin, Thomsen, Foreman, Ye
  et~al.}}]{Ludlow2008}
\bibinfo{author}{\bibfnamefont{A.~D.} \bibnamefont{Ludlow}},
  \bibinfo{author}{\bibfnamefont{T.}~\bibnamefont{Zelevinsky}},
  \bibinfo{author}{\bibfnamefont{G.~K.} \bibnamefont{Campbell}},
  \bibinfo{author}{\bibfnamefont{S.}~\bibnamefont{Blatt}},
  \bibinfo{author}{\bibfnamefont{M.~M.} \bibnamefont{Boyd}},
  \bibinfo{author}{\bibfnamefont{M.~H.~G.} \bibnamefont{de~Miranda}},
  \bibinfo{author}{\bibfnamefont{M.~J.} \bibnamefont{Martin}},
  \bibinfo{author}{\bibfnamefont{J.~W.} \bibnamefont{Thomsen}},
  \bibinfo{author}{\bibfnamefont{S.~M.} \bibnamefont{Foreman}},
  \bibinfo{author}{\bibfnamefont{J.}~\bibnamefont{Ye}}, \bibnamefont{et~al.},
  \bibinfo{journal}{Science} \textbf{\bibinfo{volume}{319}},
  \bibinfo{pages}{1805} (\bibinfo{year}{2008}).

\bibitem[{\citenamefont{Swallows et~al.}(2011)\citenamefont{Swallows, Bishof,
  Lin, Blatt, Martin, Rey, and Ye}}]{Swallows2011}
\bibinfo{author}{\bibfnamefont{M.~D.} \bibnamefont{Swallows}},
  \bibinfo{author}{\bibfnamefont{M.}~\bibnamefont{Bishof}},
  \bibinfo{author}{\bibfnamefont{Y.}~\bibnamefont{Lin}},
  \bibinfo{author}{\bibfnamefont{S.}~\bibnamefont{Blatt}},
  \bibinfo{author}{\bibfnamefont{M.~J.} \bibnamefont{Martin}},
  \bibinfo{author}{\bibfnamefont{A.~M.} \bibnamefont{Rey}}, \bibnamefont{and}
  \bibinfo{author}{\bibfnamefont{J.}~\bibnamefont{Ye}},
  \bibinfo{journal}{Science} \textbf{\bibinfo{volume}{331}},
  \bibinfo{pages}{1043} (\bibinfo{year}{2011}).

\bibitem[{\citenamefont{Maki et~al.}(1991)\citenamefont{Maki, Malcuit, Sipe,
  and Boyd}}]{Maki1991}
\bibinfo{author}{\bibfnamefont{J.~J.} \bibnamefont{Maki}},
  \bibinfo{author}{\bibfnamefont{M.~S.} \bibnamefont{Malcuit}},
  \bibinfo{author}{\bibfnamefont{J.~E.} \bibnamefont{Sipe}}, \bibnamefont{and}
  \bibinfo{author}{\bibfnamefont{R.~W.} \bibnamefont{Boyd}},
  \bibinfo{journal}{Phys. Rev. Lett.} \textbf{\bibinfo{volume}{67}},
  \bibinfo{pages}{972} (\bibinfo{year}{1991}).

\bibitem[{\citenamefont{Sautenkov et~al.}(1996)\citenamefont{Sautenkov, van
  Kampen, Eliel, and Woerdman}}]{Sautenkov1996}
\bibinfo{author}{\bibfnamefont{V.~A.} \bibnamefont{Sautenkov}},
  \bibinfo{author}{\bibfnamefont{H.}~\bibnamefont{van Kampen}},
  \bibinfo{author}{\bibfnamefont{E.~R.} \bibnamefont{Eliel}}, \bibnamefont{and}
  \bibinfo{author}{\bibfnamefont{J.~P.} \bibnamefont{Woerdman}},
  \bibinfo{journal}{Phys. Rev. Lett.} \textbf{\bibinfo{volume}{77}},
  \bibinfo{pages}{3327} (\bibinfo{year}{1996}).

\bibitem[{\citenamefont{van Kampen et~al.}(1997)\citenamefont{van Kampen,
  Sautenkov, Shalagin, Eliel, and Woerdman}}]{VanKampen1997}
\bibinfo{author}{\bibfnamefont{H.}~\bibnamefont{van Kampen}},
  \bibinfo{author}{\bibfnamefont{V.~A.} \bibnamefont{Sautenkov}},
  \bibinfo{author}{\bibfnamefont{A.~M.} \bibnamefont{Shalagin}},
  \bibinfo{author}{\bibfnamefont{E.~R.} \bibnamefont{Eliel}}, \bibnamefont{and}
  \bibinfo{author}{\bibfnamefont{J.~P.} \bibnamefont{Woerdman}},
  \bibinfo{journal}{Phys. Rev. A} \textbf{\bibinfo{volume}{56}},
  \bibinfo{pages}{3569} (\bibinfo{year}{1997}).

\bibitem[{\citenamefont{Li et~al.}(2008)\citenamefont{Li, Varzhapetyan,
  Sautenkov, Rostovtsev, Chen, Sarkisyan, and Scully}}]{Li2008}
\bibinfo{author}{\bibfnamefont{H.}~\bibnamefont{Li}},
  \bibinfo{author}{\bibfnamefont{T.}~\bibnamefont{Varzhapetyan}},
  \bibinfo{author}{\bibfnamefont{V.}~\bibnamefont{Sautenkov}},
  \bibinfo{author}{\bibfnamefont{Y.}~\bibnamefont{Rostovtsev}},
  \bibinfo{author}{\bibfnamefont{H.}~\bibnamefont{Chen}},
  \bibinfo{author}{\bibfnamefont{D.}~\bibnamefont{Sarkisyan}},
  \bibnamefont{and} \bibinfo{author}{\bibfnamefont{M.}~\bibnamefont{Scully}},
  \bibinfo{journal}{Appl. Phys. B} \textbf{\bibinfo{volume}{91}},
  \bibinfo{pages}{229} (\bibinfo{year}{2008}).

\bibitem[{\citenamefont{Li et~al.}(2009)\citenamefont{Li, Sautenkov,
  Rostovtsev, and Scully}}]{Li2009}
\bibinfo{author}{\bibfnamefont{H.}~\bibnamefont{Li}},
  \bibinfo{author}{\bibfnamefont{V.~A.} \bibnamefont{Sautenkov}},
  \bibinfo{author}{\bibfnamefont{Y.~V.} \bibnamefont{Rostovtsev}},
  \bibnamefont{and} \bibinfo{author}{\bibfnamefont{M.~O.}
  \bibnamefont{Scully}}, \bibinfo{journal}{J. Phys. B: At., Mol. Opt. Phys.}
  \textbf{\bibinfo{volume}{42}}, \bibinfo{pages}{65203} (\bibinfo{year}{2009}).

\bibitem[{\citenamefont{Shen et~al.}(2007)\citenamefont{Shen, Gao, Senin, Zhu,
  Allen, Lu, Xiao, and Eden}}]{Shen2007b}
\bibinfo{author}{\bibfnamefont{F.}~\bibnamefont{Shen}},
  \bibinfo{author}{\bibfnamefont{J.}~\bibnamefont{Gao}},
  \bibinfo{author}{\bibfnamefont{A.~A.} \bibnamefont{Senin}},
  \bibinfo{author}{\bibfnamefont{C.~J.} \bibnamefont{Zhu}},
  \bibinfo{author}{\bibfnamefont{J.~R.} \bibnamefont{Allen}},
  \bibinfo{author}{\bibfnamefont{Z.~H.} \bibnamefont{Lu}},
  \bibinfo{author}{\bibfnamefont{Y.}~\bibnamefont{Xiao}}, \bibnamefont{and}
  \bibinfo{author}{\bibfnamefont{J.~G.} \bibnamefont{Eden}},
  \bibinfo{journal}{Phys. Rev. Lett.} \textbf{\bibinfo{volume}{99}},
  \bibinfo{pages}{143201} (\bibinfo{year}{2007}).

\bibitem[{\citenamefont{Eden et~al.}(2008)\citenamefont{Eden, Ricconi, Xiao,
  Shen, and Senin}}]{Eden2008}
\bibinfo{author}{\bibfnamefont{J.}~\bibnamefont{Eden}},
  \bibinfo{author}{\bibfnamefont{B.}~\bibnamefont{Ricconi}},
  \bibinfo{author}{\bibfnamefont{Y.}~\bibnamefont{Xiao}},
  \bibinfo{author}{\bibfnamefont{F.}~\bibnamefont{Shen}}, \bibnamefont{and}
  \bibinfo{author}{\bibfnamefont{A.}~\bibnamefont{Senin}},
  \bibinfo{journal}{Adv. At., Mol., Opt. Phys.} \textbf{\bibinfo{volume}{56}},
  \bibinfo{pages}{49} (\bibinfo{year}{2008}).

\bibitem[{\citenamefont{Cundiff}(2002)}]{Cundiff2002a}
\bibinfo{author}{\bibfnamefont{S.~T.} \bibnamefont{Cundiff}},
  \bibinfo{journal}{Laser Phys.} \textbf{\bibinfo{volume}{12}},
  \bibinfo{pages}{1073} (\bibinfo{year}{2002}).

\bibitem[{\citenamefont{Lorenz and Cundiff}(2005)}]{Lorenz2005a}
\bibinfo{author}{\bibfnamefont{V.}~\bibnamefont{Lorenz}} \bibnamefont{and}
  \bibinfo{author}{\bibfnamefont{S.}~\bibnamefont{Cundiff}},
  \bibinfo{journal}{Phys. Rev. Lett.} \textbf{\bibinfo{volume}{95}},
  \bibinfo{pages}{163601} (\bibinfo{year}{2005}).

\bibitem[{\citenamefont{Lorenz et~al.}(2008)\citenamefont{Lorenz, Mukamel,
  Zhuang, and Cundiff}}]{Lorenz2008a}
\bibinfo{author}{\bibfnamefont{V.~O.} \bibnamefont{Lorenz}},
  \bibinfo{author}{\bibfnamefont{S.}~\bibnamefont{Mukamel}},
  \bibinfo{author}{\bibfnamefont{W.}~\bibnamefont{Zhuang}}, \bibnamefont{and}
  \bibinfo{author}{\bibfnamefont{S.~T.} \bibnamefont{Cundiff}},
  \bibinfo{journal}{Phys. Rev. Lett.} \textbf{\bibinfo{volume}{100}},
  \bibinfo{pages}{013603} (\bibinfo{year}{2008}).

\bibitem[{\citenamefont{Dai et~al.}(2012)\citenamefont{Dai, Richter, Li,
  Bristow, Falvo, Mukamel, and Cundiff}}]{Dai2012}
\bibinfo{author}{\bibfnamefont{X.}~\bibnamefont{Dai}},
  \bibinfo{author}{\bibfnamefont{M.}~\bibnamefont{Richter}},
  \bibinfo{author}{\bibfnamefont{H.}~\bibnamefont{Li}},
  \bibinfo{author}{\bibfnamefont{A.~D.} \bibnamefont{Bristow}},
  \bibinfo{author}{\bibfnamefont{C.}~\bibnamefont{Falvo}},
  \bibinfo{author}{\bibfnamefont{S.}~\bibnamefont{Mukamel}}, \bibnamefont{and}
  \bibinfo{author}{\bibfnamefont{S.~T.} \bibnamefont{Cundiff}},
  \bibinfo{journal}{Phys. Rev. Lett.} \textbf{\bibinfo{volume}{108}},
  \bibinfo{pages}{193201} (\bibinfo{year}{2012}).

\bibitem[{\citenamefont{Gao et~al.}(2016)\citenamefont{Gao, Cundiff, and
  Li}}]{Gao2016}
\bibinfo{author}{\bibfnamefont{F.}~\bibnamefont{Gao}},
  \bibinfo{author}{\bibfnamefont{S.~T.} \bibnamefont{Cundiff}},
  \bibnamefont{and} \bibinfo{author}{\bibfnamefont{H.}~\bibnamefont{Li}},
  \bibinfo{journal}{Opt. Lett.} \textbf{\bibinfo{volume}{41}},
  \bibinfo{pages}{2954} (\bibinfo{year}{2016}).

\bibitem[{\citenamefont{Lomsadze and Cundiff}(2018)}]{PhysRevLett.120.233401}
\bibinfo{author}{\bibfnamefont{B.}~\bibnamefont{Lomsadze}} \bibnamefont{and}
  \bibinfo{author}{\bibfnamefont{S.~T.} \bibnamefont{Cundiff}},
  \bibinfo{journal}{Phys. Rev. Lett.} \textbf{\bibinfo{volume}{120}},
  \bibinfo{pages}{233401} (\bibinfo{year}{2018}).

\bibitem[{\citenamefont{Engel et~al.}(2007)\citenamefont{Engel, Calhoun, Read,
  Ahn, Man{\v{c}}al, Cheng, Blankenship, and Fleming}}]{Engel2007}
\bibinfo{author}{\bibfnamefont{G.~S.} \bibnamefont{Engel}},
  \bibinfo{author}{\bibfnamefont{T.~R.} \bibnamefont{Calhoun}},
  \bibinfo{author}{\bibfnamefont{E.~L.} \bibnamefont{Read}},
  \bibinfo{author}{\bibfnamefont{T.-K.} \bibnamefont{Ahn}},
  \bibinfo{author}{\bibfnamefont{T.}~\bibnamefont{Man{\v{c}}al}},
  \bibinfo{author}{\bibfnamefont{Y.-C.} \bibnamefont{Cheng}},
  \bibinfo{author}{\bibfnamefont{R.~E.} \bibnamefont{Blankenship}},
  \bibnamefont{and} \bibinfo{author}{\bibfnamefont{G.~R.}
  \bibnamefont{Fleming}}, \bibinfo{journal}{Nature}
  \textbf{\bibinfo{volume}{446}}, \bibinfo{pages}{782} (\bibinfo{year}{2007}).

\bibitem[{\citenamefont{Bernien et~al.}(2017)\citenamefont{Bernien, Schwartz,
  Keesling, Levine, Omran, Pichler, Choi, Zibrov, Endres, Greiner
  et~al.}}]{Bernien2017}
\bibinfo{author}{\bibfnamefont{H.}~\bibnamefont{Bernien}},
  \bibinfo{author}{\bibfnamefont{S.}~\bibnamefont{Schwartz}},
  \bibinfo{author}{\bibfnamefont{A.}~\bibnamefont{Keesling}},
  \bibinfo{author}{\bibfnamefont{H.}~\bibnamefont{Levine}},
  \bibinfo{author}{\bibfnamefont{A.}~\bibnamefont{Omran}},
  \bibinfo{author}{\bibfnamefont{H.}~\bibnamefont{Pichler}},
  \bibinfo{author}{\bibfnamefont{S.}~\bibnamefont{Choi}},
  \bibinfo{author}{\bibfnamefont{A.~S.} \bibnamefont{Zibrov}},
  \bibinfo{author}{\bibfnamefont{M.}~\bibnamefont{Endres}},
  \bibinfo{author}{\bibfnamefont{M.}~\bibnamefont{Greiner}},
  \bibnamefont{et~al.}, \bibinfo{journal}{Nature}
  \textbf{\bibinfo{volume}{551}}, \bibinfo{pages}{579} (\bibinfo{year}{2017}).

\bibitem[{\citenamefont{Mazurenko et~al.}(2017)\citenamefont{Mazurenko, Chiu,
  Ji, Parsons, Kan{\'{a}}sz-Nagy, Schmidt, Grusdt, Demler, Greif, and
  Greiner}}]{Mazurenko2017}
\bibinfo{author}{\bibfnamefont{A.}~\bibnamefont{Mazurenko}},
  \bibinfo{author}{\bibfnamefont{C.~S.} \bibnamefont{Chiu}},
  \bibinfo{author}{\bibfnamefont{G.}~\bibnamefont{Ji}},
  \bibinfo{author}{\bibfnamefont{M.~F.} \bibnamefont{Parsons}},
  \bibinfo{author}{\bibfnamefont{M.}~\bibnamefont{Kan{\'{a}}sz-Nagy}},
  \bibinfo{author}{\bibfnamefont{R.}~\bibnamefont{Schmidt}},
  \bibinfo{author}{\bibfnamefont{F.}~\bibnamefont{Grusdt}},
  \bibinfo{author}{\bibfnamefont{E.}~\bibnamefont{Demler}},
  \bibinfo{author}{\bibfnamefont{D.}~\bibnamefont{Greif}}, \bibnamefont{and}
  \bibinfo{author}{\bibfnamefont{M.}~\bibnamefont{Greiner}},
  \bibinfo{journal}{Nature} \textbf{\bibinfo{volume}{545}},
  \bibinfo{pages}{462} (\bibinfo{year}{2017}).

\bibitem[{\citenamefont{Trotzky et~al.}(2012)\citenamefont{Trotzky, Chen,
  Flesch, McCulloch, Schollw{\"{o}}ck, Eisert, and Bloch}}]{Trotzky2012}
\bibinfo{author}{\bibfnamefont{S.}~\bibnamefont{Trotzky}},
  \bibinfo{author}{\bibfnamefont{Y.~A.} \bibnamefont{Chen}},
  \bibinfo{author}{\bibfnamefont{A.}~\bibnamefont{Flesch}},
  \bibinfo{author}{\bibfnamefont{I.~P.} \bibnamefont{McCulloch}},
  \bibinfo{author}{\bibfnamefont{U.}~\bibnamefont{Schollw{\"{o}}ck}},
  \bibinfo{author}{\bibfnamefont{J.}~\bibnamefont{Eisert}}, \bibnamefont{and}
  \bibinfo{author}{\bibfnamefont{I.}~\bibnamefont{Bloch}},
  \bibinfo{journal}{Nat. Phys.} \textbf{\bibinfo{volume}{8}},
  \bibinfo{pages}{325} (\bibinfo{year}{2012}).

\bibitem[{\citenamefont{Szudy and Baylis}(1975)}]{Szudy1975}
\bibinfo{author}{\bibfnamefont{J.}~\bibnamefont{Szudy}} \bibnamefont{and}
  \bibinfo{author}{\bibfnamefont{W.}~\bibnamefont{Baylis}},
  \bibinfo{journal}{J. Quant. Spectrosc. Radiat. Transfer}
  \textbf{\bibinfo{volume}{15}}, \bibinfo{pages}{641} (\bibinfo{year}{1975}).

\bibitem[{\citenamefont{Reuven}(1975)}]{Reuven1975}
\bibinfo{author}{\bibfnamefont{A.}~\bibnamefont{Reuven}},
  \bibinfo{journal}{Adv. Chem. Phys.} \textbf{\bibinfo{volume}{33}},
  \bibinfo{pages}{235} (\bibinfo{year}{1975}).

\bibitem[{\citenamefont{Smith et~al.}(1973)\citenamefont{Smith, Cooper, and
  Roszman}}]{Smith1973}
\bibinfo{author}{\bibfnamefont{E.~W.} \bibnamefont{Smith}},
  \bibinfo{author}{\bibfnamefont{J.}~\bibnamefont{Cooper}}, \bibnamefont{and}
  \bibinfo{author}{\bibfnamefont{L.~J.} \bibnamefont{Roszman}},
  \bibinfo{journal}{J. Quant. Spectrosc. Radiat. Transfer}
  \textbf{\bibinfo{volume}{13}}, \bibinfo{pages}{1523} (\bibinfo{year}{1973}).

\bibitem[{\citenamefont{Ali and Griem}(1965)}]{Ali1965}
\bibinfo{author}{\bibfnamefont{A.~W.} \bibnamefont{Ali}} \bibnamefont{and}
  \bibinfo{author}{\bibfnamefont{H.~R.} \bibnamefont{Griem}},
  \bibinfo{journal}{Phys. Rev.} \textbf{\bibinfo{volume}{140}},
  \bibinfo{pages}{A1044} (\bibinfo{year}{1965}).

\bibitem[{\citenamefont{Anderson}(1949)}]{Anderson1949}
\bibinfo{author}{\bibfnamefont{P.~W.} \bibnamefont{Anderson}},
  \bibinfo{journal}{Phys. Rev.} \textbf{\bibinfo{volume}{76}},
  \bibinfo{pages}{647} (\bibinfo{year}{1949}).

\bibitem[{\citenamefont{Lewis}(1980)}]{Lewis1980}
\bibinfo{author}{\bibfnamefont{E.}~\bibnamefont{Lewis}},
  \bibinfo{journal}{Phys. Rep.} \textbf{\bibinfo{volume}{58}},
  \bibinfo{pages}{1} (\bibinfo{year}{1980}).

\bibitem[{\citenamefont{Weisskopf}(1932)}]{Weisskopf1932}
\bibinfo{author}{\bibfnamefont{V.}~\bibnamefont{Weisskopf}},
  \bibinfo{journal}{Zeitschrift f{\"u}r Physik} \textbf{\bibinfo{volume}{75}},
  \bibinfo{pages}{287} (\bibinfo{year}{1932}).

\bibitem[{\citenamefont{Thorne et~al.}(1999)\citenamefont{Thorne, Litzén,
  Johansson, and Litzen}}]{Thorne1999}
\bibinfo{author}{\bibfnamefont{A.}~\bibnamefont{Thorne}},
  \bibinfo{author}{\bibfnamefont{U.}~\bibnamefont{Litzén}},
  \bibinfo{author}{\bibfnamefont{S.}~\bibnamefont{Johansson}},
  \bibnamefont{and} \bibinfo{author}{\bibfnamefont{U.}~\bibnamefont{Litzen}},
  \emph{\bibinfo{title}{Spectrophysics: Principles and Applications}}
  (\bibinfo{publisher}{Springer}, \bibinfo{year}{1999}).

\bibitem[{\citenamefont{Leegwater and Mukamel}(1994)}]{Leegwater1994a}
\bibinfo{author}{\bibfnamefont{J.~A.} \bibnamefont{Leegwater}}
  \bibnamefont{and} \bibinfo{author}{\bibfnamefont{S.}~\bibnamefont{Mukamel}},
  \bibinfo{journal}{Phys. Rev. A} \textbf{\bibinfo{volume}{49}},
  \bibinfo{pages}{146} (\bibinfo{year}{1994}).

\bibitem[{\citenamefont{Nardin et~al.}(2013)\citenamefont{Nardin, Autry,
  Silverman, and Cundiff}}]{Nardin2013}
\bibinfo{author}{\bibfnamefont{G.}~\bibnamefont{Nardin}},
  \bibinfo{author}{\bibfnamefont{T.~M.} \bibnamefont{Autry}},
  \bibinfo{author}{\bibfnamefont{K.~L.} \bibnamefont{Silverman}},
  \bibnamefont{and} \bibinfo{author}{\bibfnamefont{S.~T.}
  \bibnamefont{Cundiff}}, \bibinfo{journal}{Opt. Express}
  \textbf{\bibinfo{volume}{21}}, \bibinfo{pages}{28617} (\bibinfo{year}{2013}).

\bibitem[{\citenamefont{Tekavec et~al.}(2007)\citenamefont{Tekavec, Lott, and
  Marcus}}]{Tekavec2007}
\bibinfo{author}{\bibfnamefont{P.~F.} \bibnamefont{Tekavec}},
  \bibinfo{author}{\bibfnamefont{G.~A.} \bibnamefont{Lott}}, \bibnamefont{and}
  \bibinfo{author}{\bibfnamefont{A.~H.} \bibnamefont{Marcus}},
  \bibinfo{journal}{J. Chem. Phys.} \textbf{\bibinfo{volume}{127}},
  \bibinfo{pages}{214307} (\bibinfo{year}{2007}).

\bibitem[{\citenamefont{Bruder et~al.}(2015)\citenamefont{Bruder, Binz, and
  Stienkemeier}}]{Bruder2015}
\bibinfo{author}{\bibfnamefont{L.}~\bibnamefont{Bruder}},
  \bibinfo{author}{\bibfnamefont{M.}~\bibnamefont{Binz}}, \bibnamefont{and}
  \bibinfo{author}{\bibfnamefont{F.}~\bibnamefont{Stienkemeier}},
  \bibinfo{journal}{Phys. Rev. A} \textbf{\bibinfo{volume}{92}}
  (\bibinfo{year}{2015}).

\bibitem[{\citenamefont{Yu et~al.}(2018)\citenamefont{Yu, Titze, Zhu, Liu, and
  Li}}]{Yu2018}
\bibinfo{author}{\bibfnamefont{S.}~\bibnamefont{Yu}},
  \bibinfo{author}{\bibfnamefont{M.}~\bibnamefont{Titze}},
  \bibinfo{author}{\bibfnamefont{Y.}~\bibnamefont{Zhu}},
  \bibinfo{author}{\bibfnamefont{X.}~\bibnamefont{Liu}}, \bibnamefont{and}
  \bibinfo{author}{\bibfnamefont{H.}~\bibnamefont{Li}}, \bibinfo{journal}{arXiv
  preprint arXiv:1807.09300}  (\bibinfo{year}{2018}).

\bibitem[{\citenamefont{Nesmeyanov}(1963)}]{Nesmeyanov1963}
\bibinfo{author}{\bibfnamefont{A.~N.} \bibnamefont{Nesmeyanov}},
  \emph{\bibinfo{title}{Vapor pressure curve of chemical elements}}
  (\bibinfo{publisher}{Elsevier, New York}, \bibinfo{year}{1963}).

\bibitem[{\citenamefont{Cline et~al.}(1994)\citenamefont{Cline, Miller, and
  Heinzen}}]{Cline1994}
\bibinfo{author}{\bibfnamefont{R.~A.} \bibnamefont{Cline}},
  \bibinfo{author}{\bibfnamefont{J.~D.} \bibnamefont{Miller}},
  \bibnamefont{and} \bibinfo{author}{\bibfnamefont{D.~J.}
  \bibnamefont{Heinzen}}, \bibinfo{journal}{Phys. Rev. Lett.}
  \textbf{\bibinfo{volume}{73}}, \bibinfo{pages}{632} (\bibinfo{year}{1994}).

\bibitem[{\citenamefont{Lomsadze and Cundiff}(2017)}]{Lomsadze2017b}
\bibinfo{author}{\bibfnamefont{B.}~\bibnamefont{Lomsadze}} \bibnamefont{and}
  \bibinfo{author}{\bibfnamefont{S.~T.} \bibnamefont{Cundiff}},
  \bibinfo{journal}{Science} \textbf{\bibinfo{volume}{357}},
  \bibinfo{pages}{1389} (\bibinfo{year}{2017}).

\end{thebibliography}






\end{document}